\documentclass[11pt, epsfig]{article}
\usepackage{epsfig, amsmath, amssymb, amsthm, times}
\usepackage{graphicx}
\DeclareMathOperator{\esssup}{ess\,sup}

\newtheorem {thm}{Theorem}[section]
\newtheorem {lem}[thm]{Lemma}

\newtheorem {cor}[thm]{Corollary}

\theoremstyle{defintion}
\newtheorem {df}[thm]{Definition}

\theoremstyle{remark}
\newtheorem{rem}[thm]{Remark}

\theoremstyle{example}
\newtheorem{ex}[thm]{Example}

\theoremstyle{assumption}

\def\pf{{\it Proof.\;}}
\def\V{\mathrm{V}}

\def\E{{\mathbb E}}
\def\P{{\mathbb P}}
\def\R{{\mathbb R}}
\def\N{{\mathbb N}}

\def\Z{{\mathbb Z}}
\def\SS{{\mathbb S}}

\def\lbl{\label}
\def\be{\begin{equation}}
\def\ee{\end{equation}}
\def\p{\partial}
\def\qed{\square}
\def\t{\intercal}

\title{The nonlinear heat equation on $W$-random graphs
}
\author{Georgi S. Medvedev 
\thanks{
Department of Mathematics, Drexel University, 3141 Chestnut Street,
Philadelphia, PA 19104, {\tt medvedev@drexel.edu} 
}
}

\begin{document}
\maketitle
\begin{abstract}
For systems of coupled differential equations on a  sequence of $W$-random graphs,
we derive  the continuum limit in the form of an evolution integral equation.
We prove that solutions of the initial value problems (IVPs) for the discrete model converge
to the solution of the IVP for its continuum limit. These results combined with the analysis
of nonlocally coupled deterministic networks in \cite{GKVM13} justify the continuum
(thermodynamic) limit for a large class of coupled dynamical systems on convergent
families of graphs. 
\end{abstract}

\section{Introduction}
In this paper, we study coupled dynamical systems on a sequence of
graphs $\{G_n\}$:
\be\lbl{in.W}
{d\over dt} u_{ni}(t) =n^{-1} \sum_{j: (i,j)\in E( G_n)}
D(u_{nj}-u_{ni}), \quad i\in [n].
\ee
Here, $G_n=\langle [n], E(G_n)\rangle$ is a graph on $n$ nodes and 
$D$ is a Lipschitz continuous function. The operator on the right-hand
side of (\ref{in.W}) models the nonlinear diffusion across edges of $G_n$.
Thus, we refer to (\ref{in.W}) as a nonlinear heat equation on $G_n$.

The evolution equations like (\ref{in.W}) are used in modeling diverse systems
ranging from neuronal networks in biology \cite{HopIzh-book, StrSync, ErmKop91, MZ12}, 
to Josephson junctions and coupled lasers 
in physics \cite{LiErn92, PhiZan93}, to communication, sensor, and power
networks in technology \cite{DorBul12, Med12}. The Kuramoto model,
a prominent example of (\ref{in.W}), is widely used as a paradigm for 
studying collective dynamics of coupled oscillators of diverse nature
\cite{Kur84, Kur84-book, GirHas12, DorBul12, WilStr06}.

 In this paper, we are interested in the case when $\{G_n\}$ is a sequence
of dense graphs, i.e., $|E(G_n)|=O(n^2)$.  This corresponds to the 
nonlocal diffusion operator in (\ref{in.W}). Nonlocally coupled systems
have attracted much attention in nonlinear science recently
\cite{KurBat02,TanKur03, ShiKur04, OttAnt08, WatStr98,
  WilStr06,GirHas12}. They arise as  models of diverse phenomena
throughout physics
and biology; and feature several remarkable effects, such as  
chimera states and coherence-incoherence transition
(see, e.g., \cite{KurBat02, ShiKur04, Kur95, Lai09,
 OmeMaiTas08, OmeHov11, OmeRie12, OmeWol10, OmeMai10}).
Overall, nonlocally coupled dynamical systems are less understood
than systems with local coupling.

\begin{figure}
{\bf a}\hspace{0.1 cm}\includegraphics[height=1.8in,width=2.0in]{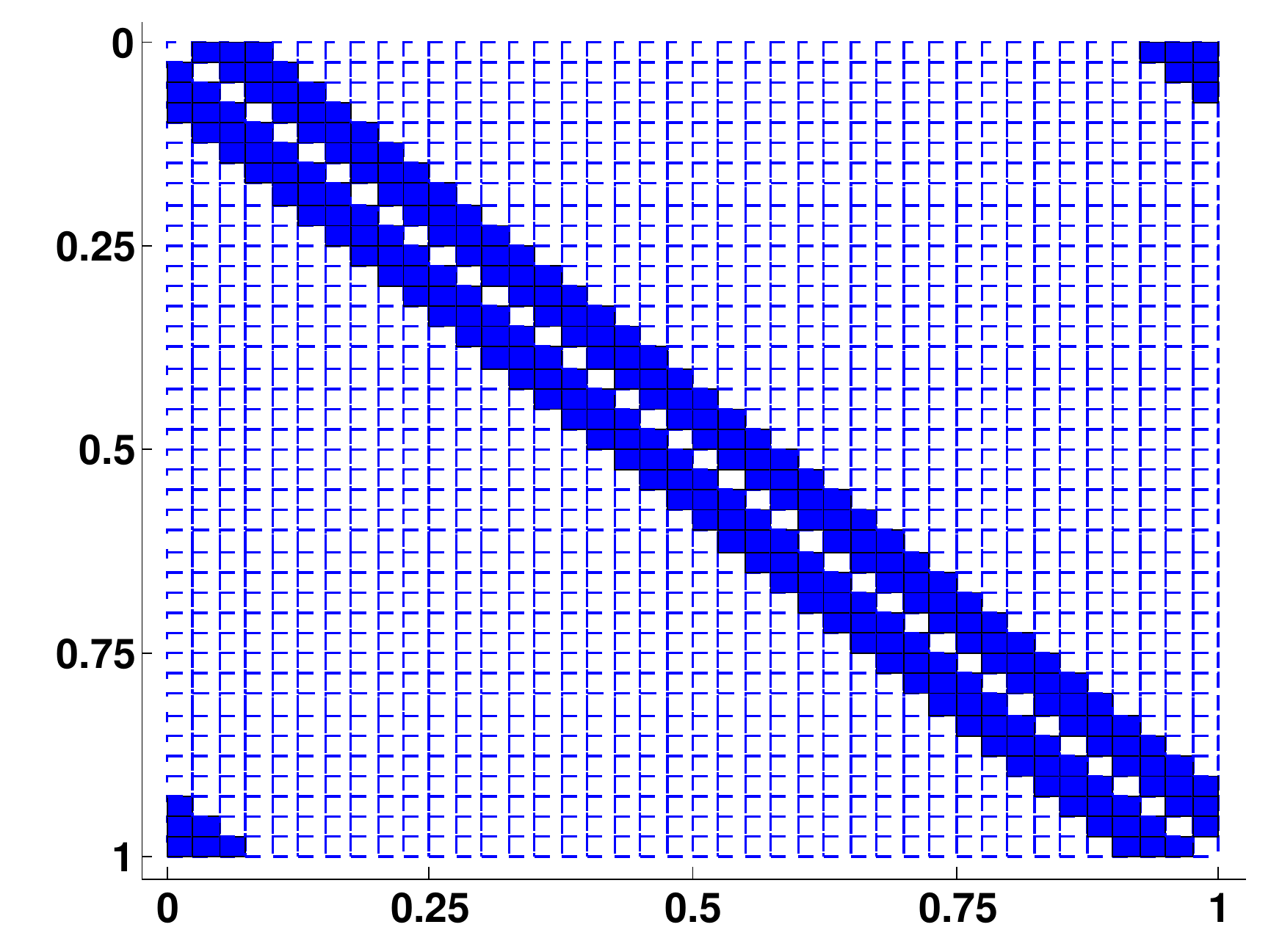}\hspace{0.1 cm}
{\bf b}\hspace{0.1 cm}\includegraphics[height=1.8in,width=2.0in]{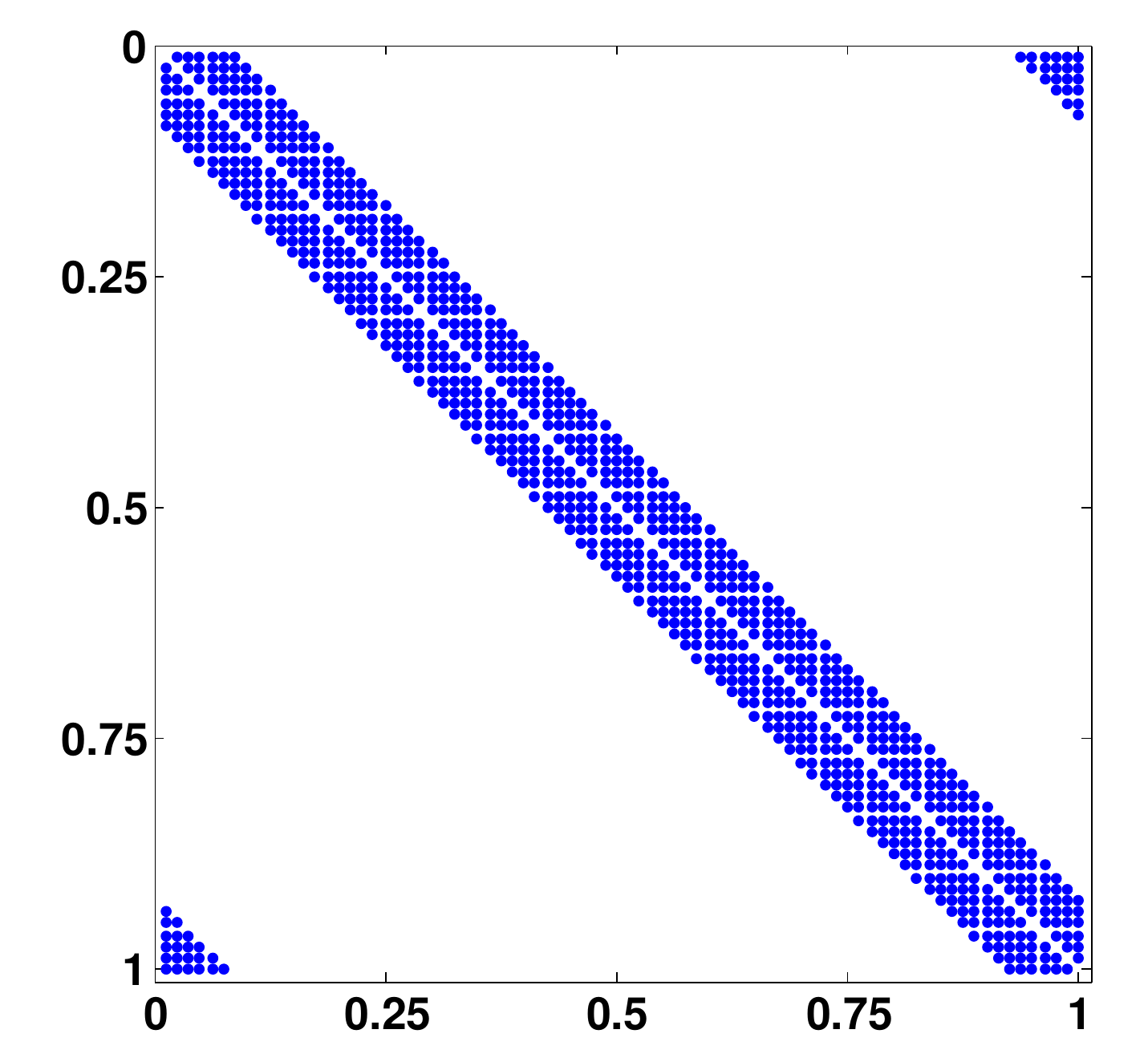}\hspace{0.1 cm}
{\bf c}\hspace{0.1 cm}\includegraphics[height=1.8in,width=2.0in]{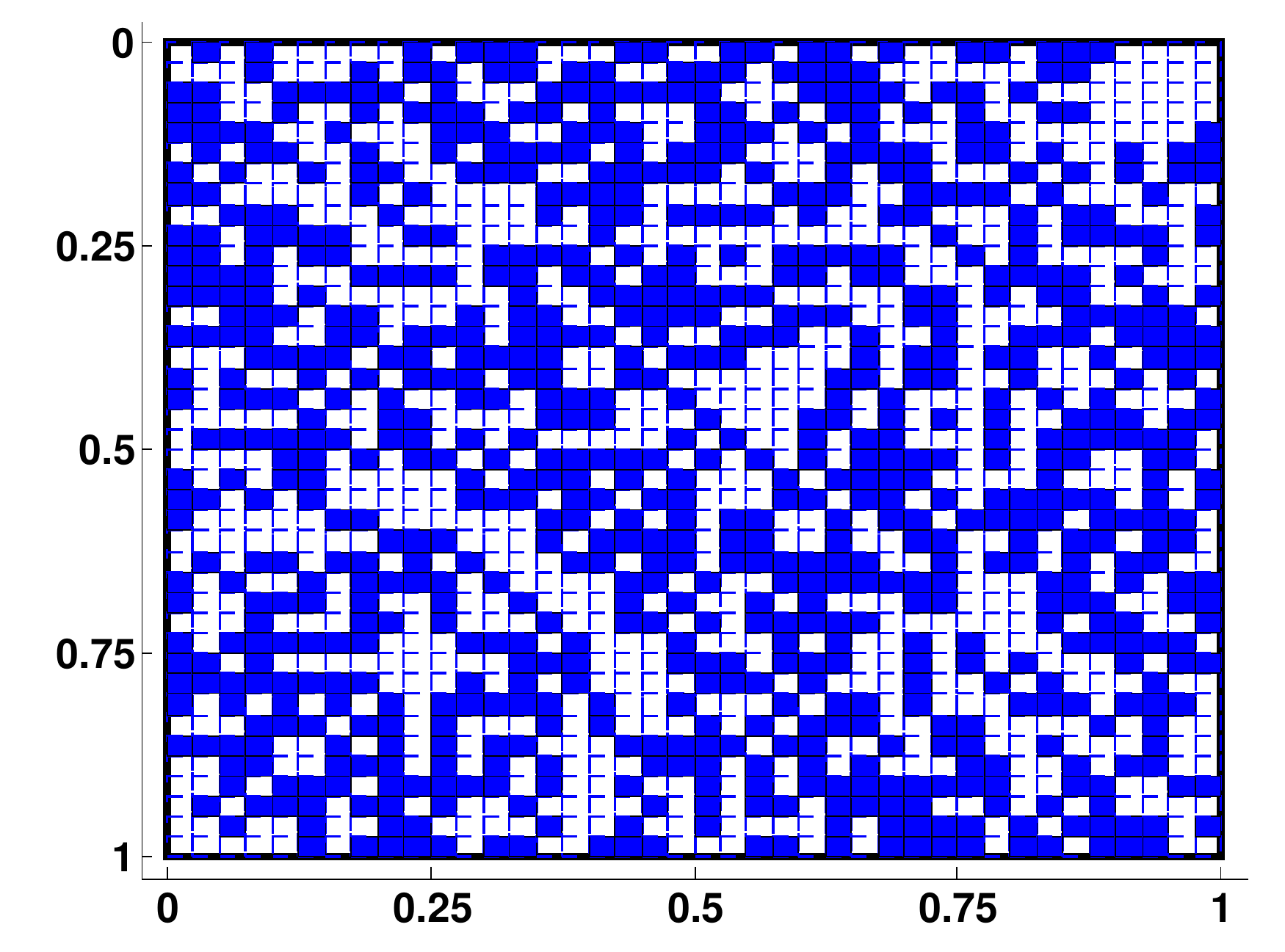}
\caption{ The plot of the support of $W_{G_n}$ ({\bf a}) and that of the support of its limit $W_G$ ({\bf b}).
{\bf c} The plot of the support of $W_{G(n,p)}$. Each function is defined on
a unit square and is equal to $1$ on cells colored in black and $0$
otherwise. The direction of the $y$-axis was chosen to emphasize the relation of these functions
to the adjacency matrices of the corresponding graphs.
}
\lbl{f.1}
\end{figure}

For analyzing nonlocally coupled systems, the continuum (thermodynamic) limit
proved to be a very useful tool \cite{KurBat02, OttAnt08,WilStr06,GirHas12}.
As $n\to\infty$, one can formally interpret the right-hand side of (\ref{in.W}) 
as a Riemann sum to obtain
\be\lbl{in.Wcont}
{\p\over\p t} u(x,t)=\int_I W(x,y) D\left( u(y,t)-u(x,t)\right) dy,
\ee
where $u(x,t)$ now describes a continuum of (local) dynamical systems
distributed along $I:=[0,1]$. For some patterns of connectivity, the kernel
$W$ in (\ref{in.Wcont}) can be guessed from the pixel picture of the 
adjacency matrix of $G_n$ \cite{BorChay11, LovGraphLim12}. 
For example, let $G_n$ be a graph on $n$ nodes
distributed uniformly along  a circle, and let $k=\lfloor rn\rfloor$ for fixed
$r\in (0,1)$. Suppose each node of $G_n$ is connected to $k$ of its nearest
neighbors from each side, i.e., $G_n$ is a $k$-nearest-neighbor graph.
The pixel picture of $G_n$ is shown in Figure~\ref{f.1}a. Specifically, 
Figure~\ref{f.1}a shows the support of the $\{0,1\}$-valued function
$W_{G_n}:~[0,1]^2\to\{0,1\}$ such that
$$
W_{G_n}(x,y)=1\;\mbox{if}\; (i,j)\in E(G_n)\;\mbox{and}\; 
(x,y)\in [(i-1)n^{-1}, in^{-1})\times [(j-1)n^{-1}, jn^{-1}), \;(i,j)\in [n]^2.
$$
Function $W_{G_n}$ provides the geometric representation of the adjacency matrix
of $G_n$.
It is easy to see
that as $n\to\infty$, $\{W_{G_n}\}$ converges  to the $\{0,1\}$-valued function,
whose support is shown in Fig.~\ref{f.1}b.
This is the limit of the $k$-nearest-neighbor family of graphs $\{G_n\}$.
A less obvious example is shown in Figure~\ref{f.1}c. Here, we show a pixel picture for 
the Erd\H{o}s-R\'{e}nyi graph $G(n,0.5)$, which converges to the constant function $0.5$ as 
$n\to\infty$ (cf.~\cite{LovGraphLim12}).

The formally derived continuum limit (\ref{in.Wcont}) was used to study the discrete model
(\ref{in.W}) for large $n$ in many papers \cite{KurBat02, OttAnt08, WatStr98,WilStr06,GirHas12}. 
In \cite{GKVM13}, Grinshpan, Kaliuzhnyi-Verbovetskyi, and the author
 provided a rigorous justification of the continuum limit (\ref{in.Wcont}). 
The analysis of the continuum limit in \cite{GKVM13} 
uses the ideas from the theory of 
graph limits \cite{LovGraphLim12,LovSze06,BorChay08},
which for every convergent family  of dense graphs defines the limiting object, a 
measurable symmetric function $W$. This function is called a graphon.
It captures the connectivity of $G_n$ for large $n$. In \cite{GKVM13}, for convergent
sequences of deterministic graphs $\{G_n\}$, it was shown that 
with the kernel of the integral operator on the right-hand side of (\ref{in.Wcont}) taken to be 
the limit of $\{G_n\}$, the solution of the IVP
for (\ref{in.Wcont}) approximates those of the IVPs for (\ref{in.W}) for large $n$. 

The analysis in \cite{GKVM13} does not cover dynamical systems on random graphs.
The latter have many important applications \cite{WatStr98, WilStr06}. Thus, in this paper, 
we focus on systems on random graphs. Specifically, we prove  convergence of solutions of the IVPs for
(\ref{in.W}) on $W$-random graphs $G_n$ to the solution of the IVP for (\ref{in.Wcont}).
A $W$-random graph is constructed from a graphon $W$ \cite{LovSze06,LovGraphLim12}. 
This construction provides a convenient general analytical model for random graphs, which 
includes many random graphs that are 
important in applications, such as Erd\H{o}s-R\'{e}nyi and small-world (SW) graphs 
(see Figs.~\ref{f.1}c and \ref{f.2}b,c)
\cite{Boll-RandGraph, JanLuc-RandGraph, WatStr98}. At the same time, $W$-random
graphs fit naturally into the convergence analysis of the families of discrete models
like (\ref{in.W}).

The remainder of this paper is organized as follows. In the next section, we formulate 
the IVPs for the discrete model and its continuum limit. In Sections~\ref{sec.rand}
and \ref{sec.det}, we prove convergence of solutions of discrete models for
two different variants of $W$-random graphs. In the variant, analyzed in Section~\ref{sec.rand},
the right-hand side of ({\ref{in.W}) can be interpreted as the Monte-Carlo approximation
of the integral on the right-hand side of (\ref{in.Wcont}). Consistent with this interpretation,
we find that the rate of convergence of the solutions of discrete problems (in $C(0,T;L^2(I))$
norm) is $O(n^{-1/2})$. In the variant of the random network model considered in Section~\ref{sec.det}, 
which was included for
the sake of convenience in applications, the rate of convergence also depends 
on the regularity of the graphon $W$.   As an application of our results, in Section~\ref{sec.small}
we derive the continuum limit for dynamical systems on SW graphs \cite{WatStr98, WilStr06}
(see Fig.~\ref{f.2}).
We conclude with the discussion of our results in Section~\ref{sec.final}. 
\begin{figure}
{\bf a}\includegraphics[height=1.8in,width=2.0in]{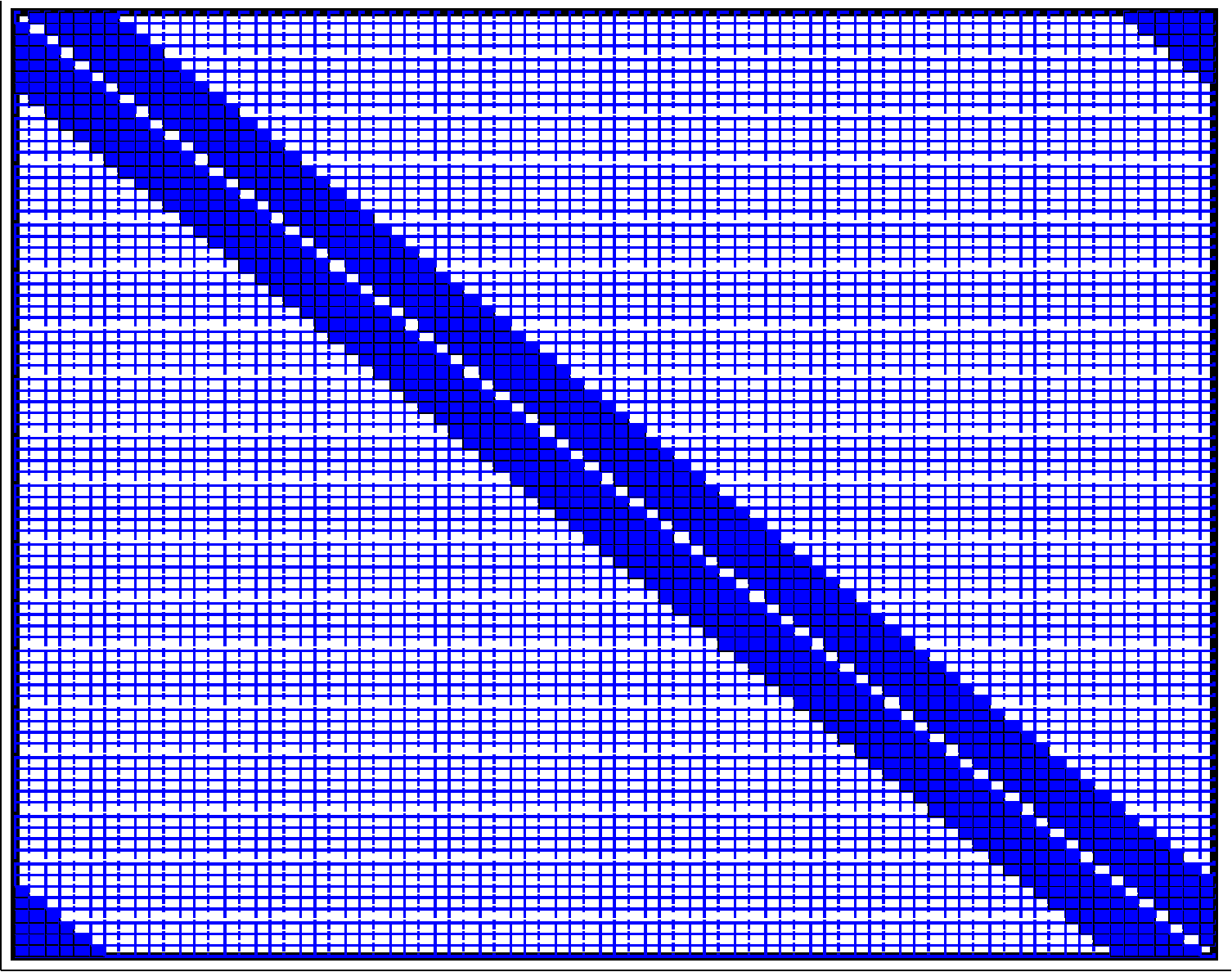}
{\bf b}\includegraphics[height=1.8in,width=2.0in]{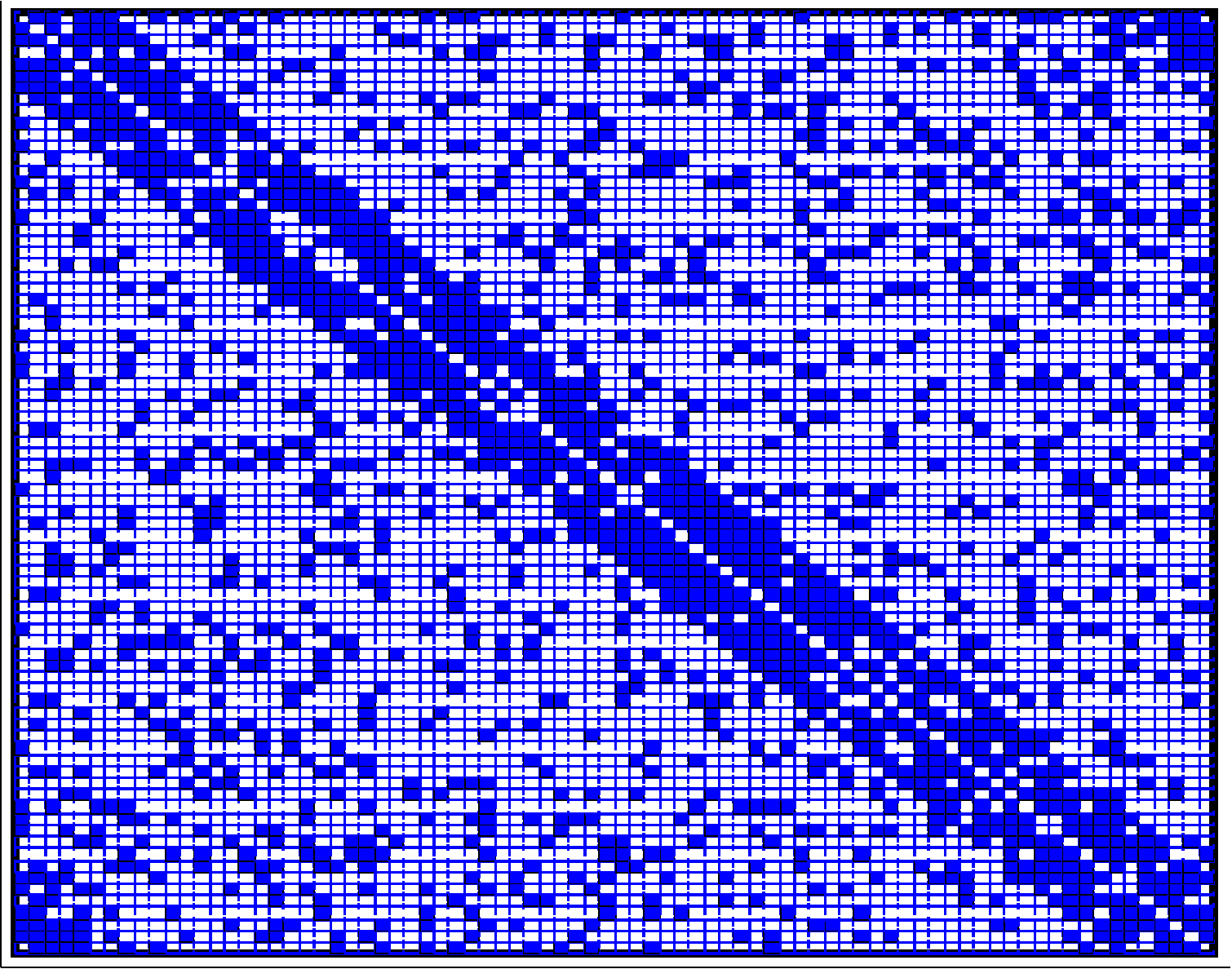}
{\bf c}\includegraphics[height=1.8in,width=2.0in]{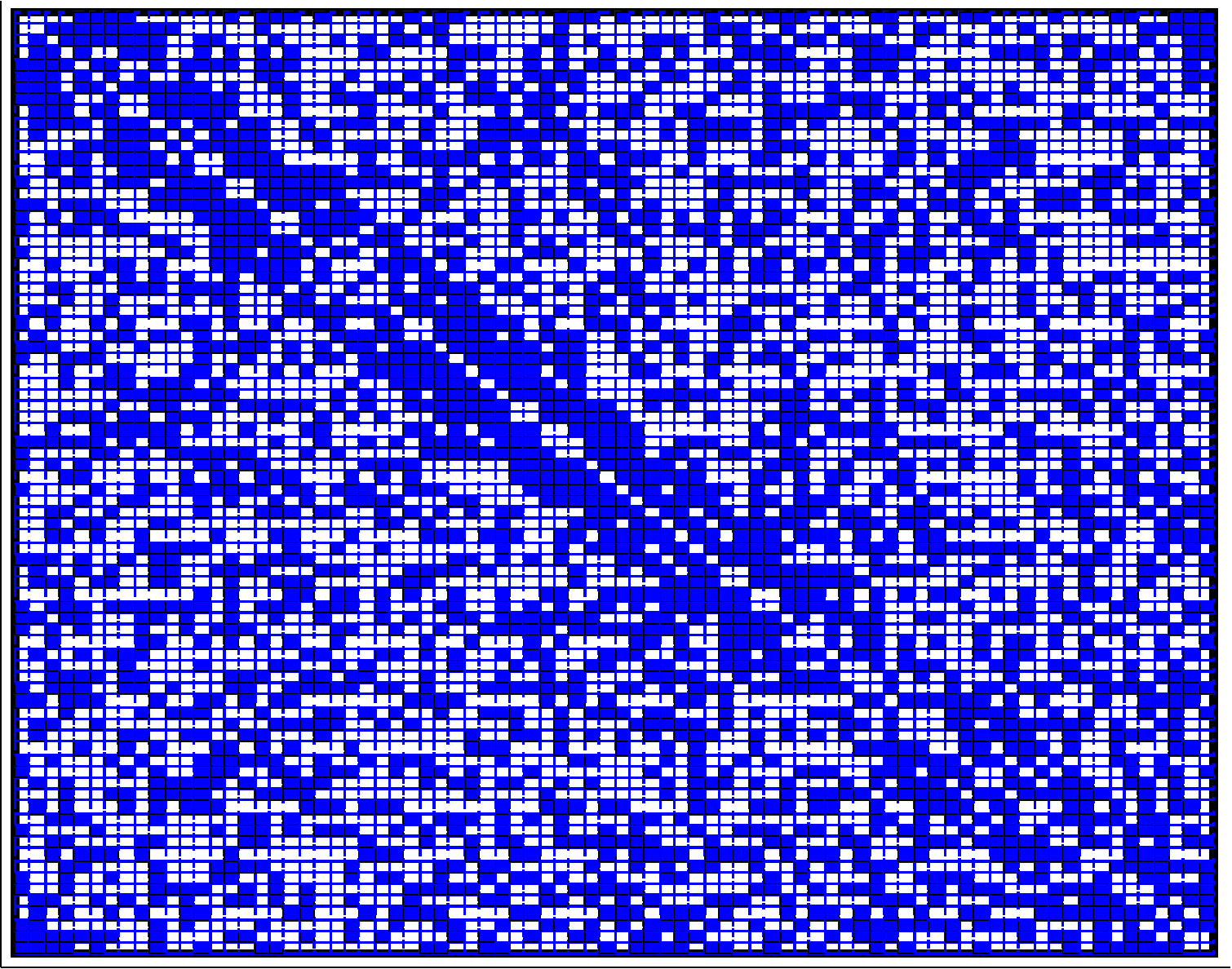}
\caption{The pixel pictures of the $k-$nearest-neighbor network on 
a ring (a) and two small-world graphs (b,c) that were obtained from
the network in (a) by replacing local connections with random long-range
connections.
}
\lbl{f.2}
\end{figure}

\section{The discrete model and its continuum limit}\lbl{sec.ivps}
\setcounter{equation}{0}
Throughout this paper, we assume that $W(x,y)$ belongs to $\mathcal{W}_0$, 
a class of  symmetric measurable functions on $I^2$ with values in $[0,1]$. 
$W$ represents the limit of a convergent family of dense graphs 
$\{G_n\}$ (see \cite{LovGraphLim12}, for an exposition of the theory of graph limits;
see also Section 2 in \cite{GKVM13} for a brief review of facts
from this theory that are relevant for constructing continuum limits of dynamical networks.)

Let  $X_n=\{x_{n1}, x_{n2},\dots, x_{nn}\}$ be 
a set of distinct points from $I$ 
and $W\in\mathcal{W}_0$.
In this section, we introduce IVPs for the nonlinear heat equation on
$G_n=\langle V(G_n), E(G_n)\rangle$, a certain graph on $n$ nodes, 
constructed using $W$ and $X_n$.

The sequence of graphs $\{G_n\}$ will be defined below. Suppose $G_n$ is given.
By the IVP for the nonlinear heat equation on $G_n$, we mean 
\begin{eqnarray}\lbl{Wheat}
{d\over dt} u_{ni}(t) &=& n^{-1} \sum_{j: (i,j)\in E( G_n)} D(u_{nj}-u_{ni}),\\
\lbl{Wheat-ic}
u_{ni}(0)&=&g(x_i),\; i\in [n],
\end{eqnarray}
where $u_n(t)=(u_{n1}(t), u_{n2}(t), \dots,u_{nn}(t))$ is the unknown function. 
Here, $D(\cdot)$ is a Lipschitz function on $\R$ and $g$ is a bounded
measurable function on $I$.

The solution of the IVP for the  discrete model
(\ref{Wheat}), (\ref{Wheat-ic}) will be compared   with the solution
of the IVP for the continuum limit 
\begin{eqnarray}\lbl{Wcont}
{\p\over\p t} u(x,t)&=&\int_I W(x,y) D\left( u(y,t)-u(x,t)\right) dy,\\
\lbl{Wcont-ic}
u(x,0) &=& g(x), \, x\in I.
\end{eqnarray}
For $W\in\mathcal{W}_0$, $g\in L^\infty(I)$, and a Lipschitz continuous $D$,
there is a unique strong solution of (\ref{Wcont}), (\ref{Wcont-ic}) 
$\mathbf{u}\in C^1(\R;L^\infty(I))$ \cite{GKVM13}.
Here and below, we use bold font to denote vector-valued functions, e.g., 
$\mathbf{u}(t)=u(\cdot,t)\in L^\infty(I)$. 

Denote the projection of the solution of the continuous problem (\ref{Wcont}),
(\ref{Wcont-ic}), $u(x,t)$ onto $X_n$ by
$$
\mathbf{P}_{X_n} u(x,t)= (u(x_{n1}, t), u(x_{n2},t),\dots, u_n(x_{nn},t)).
$$

Both functions $u_n(t)$ and $\mathbf{P}_{X_n} u(x,t)$ are defined  on the discrete set $X_n$.
For such functions, we will use the weighted Euclidean inner product
$$
(u,v)_n={1\over n} \sum_{i=1}^n u_iv_i, \; u=(u_1,u_2,\dots, u_n)^\t, \; 
v=(v_1, v_2, \dots, v_n)^\t
$$ 
and the corresponding norm $\|u\|_{2,n}=\sqrt{(u,u)_n}$. 
Below, we will
use $\|\cdot\|_{2,n}$ to study the difference between the 
solutions of the discrete and continuous problems  (\ref{Wheat}) and (\ref{Wcont}) on 
$W-$random graphs.


\section{Networks on W-random graphs generated by random sequences}
\lbl{sec.rand}
\setcounter{equation}{0}
 
Denote
\be\lbl{set-tildeXn}
\tilde X=(x_1, x_2, x_3, \dots)\;\mbox{and}\; \tilde X_n=(x_1,x_2,\dots, x_n),
\ee
where $x_i, i\in\N$ are independent identically distributed (IID) random variables (RVs).
RV $x_1$ has uniform on $I$ distribution, i.e., $\mathcal{L}(x_1)=U(I)$.

\begin{df} (cf.~\cite{LovSze06})
 By a $W-$random graph on $n$ nodes generated by the random
  sequence $\tilde X$, denoted $\tilde G_n=\mathbb{G}(\tilde X_n, W)$, we mean
$\tilde G_n=\langle [n], E(\tilde G_n)\rangle$ such that
 the edges of $\tilde G_n$ are selected at random   and 
$$
\P\left\{ (i,j)\in E(\tilde G_n)\right\} = W(x_i, x_j), \;\mbox{for
  each}\; (i,j)\in [n]^2,\; i\neq j.
$$    
The decision whether to include a pair $(i,j)\in [n]^2,\; i\neq j,$ is  made independently from the
decisions for other pairs.
\end{df}
\begin{rem}
The graph sequence $\{\tilde G_n\}$ converges to graphon $W$ almost surely as 
$n\to\infty$ \cite{LovSze06}.
\end{rem}

\begin{thm}\lbl{thm.W-convergence}
Suppose $W\in\mathcal{W}_0$, $D$ is a Lipschitz continuous 
function on $\R$, and $g\in L^\infty(I)$. Let $T>0$ and suppose that
the solution of the IVP (\ref{Wcont}) and (\ref{Wcont-ic}) $u(x,t)$
satisfies the following inequality
\be\lbl{new-assume}
\min_{t\in [0,T]} \int_I \left\{ 
\int_I W(x,y) D\left(u(y,t)-u(x,t)\right)^2dy -  
\left( \int_I W(x,y) D\left(u(y,t)-u(x,t)\right)dy\right)^2\right\} dx\ge C_1
\ee
for some positive constant $C_1$.
Then  the solutions of the IVPs for the
discrete and continuum models (\ref{Wheat}), (\ref{Wheat-ic}) and 
(\ref{Wcont}), (\ref{Wcont-ic}) satisfy the following relation
$$
\lim_{n\to\infty}
\P \left\{ n^{1/2}\sup_{t\in [0,T]} \|u^{(n)}(t)-\mathbf{P}_{\tilde X_n}
u(x,t)\|_{2,n} \le C
\right\}=1
$$
for some constant $C >0$. 
\end{thm}
\begin{rem} The integral expression in (\ref{new-assume}) defines a continuous
function of $t$. This follows from $\mathbf{u}\in C(0,T;L^\infty(I))$, $\|W\|_{L^\infty(I^2)=1}$,
and Lipschitz continuity of $D$. This justifies the use of $\min$ in (\ref{new-assume}).
\end{rem}

For the proof of this theorem we will need the following application of the Central Limit Theorem
(CLT) \cite{Bill-Prob}.
\begin{lem}\lbl{lem.clt}
Suppose $W\in\mathcal{W}_0,$  $f\in L^\infty(I^2),$ and
$$
X=(x_1, x_2, x_3, \dots),
$$
where $x_i, i\in\N,$ are  IID RVs with $\mathcal{L}(x_1)=U(I)$. 
 Define RVs $\{\xi_{ij}\}, (i,j)\in\N^2,$ such that
$\mathcal{L}\left(\xi_{ij} |X\right)=\mbox{Bin}\left(W(x_i,x_j)\right).$
\footnote{$\mbox{Bin}\left( p\right)$ stands for the binomial
distribution with parameter $p\in [0,1]$.} Specifically,
\be\lbl{define-xiij}
\P\left(\xi_{ij}=1|X\right)= W(x_i,x_j)\;\mbox{and}\;
\P\left(\xi_{ij}=0|X\right)= 1-W(x_i,x_j).
\ee

Further, let
\be\lbl{etaij}
\eta_{ij}=\xi_{ij} f(x_i,x_j), \; (i,j)\in \N^2,
\ee
\be\lbl{define-zi}
z_{ni}={1\over n}\sum_{j=1}^n \eta_{ij} - \int_I f(x_i,y) W(x_i,y)dy, \quad
\mbox{and}\quad S_n=\sum_{i=1}^n z_{ni}^2.
\ee
Finally, we assume 
\be\lbl{1-positive-sigma}
\sigma^2:=\int_{I^2} f(x, y)^2 W(x,y) dxdy -\int_{I}\left(\int_{I} f(x, y)
  W(x,y) dy\right)^2 dx>0.
\ee
Then 
\be\lbl{clt}
{ S_n-\sigma^2 \over  n^{-1/2}\sqrt{ 5\sigma^4+O(n^{-1})}   }
\overset{d}{\longrightarrow} \mathcal{N}(0,1),
\ee
where $\overset{d}{\longrightarrow}$ denotes convergence in distribution, and
$\mathcal{N}(0,1)$ stands for the standard normal distribution.
\end{lem}
By construction, $\{\eta_{ij}\}$ are IID RVs. Moreover, from (\ref{define-xiij}) and (\ref{etaij})
we have 
\be\lbl{Eeta-cond}
\mu(x_i)=\E\left(\eta_{ij}|x_i\right)=\int_I f(x_i, y) W(x_i,y) dy.
\ee
Therefore, 
\begin{eqnarray}\lbl{Eeta}
\mu:=\E\eta_{ij} &=& \E\E\left(\eta_{ij}|x_i\right)=\int_{I^2} f(x, y) W(x,y) dxdy,\\
\nonumber
\V\eta_{ij} & = &
\E\E \left(\left( \eta_{ij} -\mu \right)^2|x_i\right)=
\E\E\left((\eta_{ij}^2|x_i)-2\mu \E(\eta_{ij}|x_i)+\mu^2\right)\\
\lbl{Veta}     &=& \int_{I^2} f(x, y)^2 W(x,y) dxdy - \int_{I} \left(\int_{I} f(x, y) W(x,y) dy\right)^2 dx
=\sigma^2.
\end{eqnarray}

Let 
\be\lbl{def-yni}
y_{ni}=\sqrt{n}z_{ni}.
\ee
We prove  (\ref{clt}) by applying the CLT to $\sum_{i=1}^n y_{ni}^2$.
To justify the application of the CLT,
we need to compute  three first  moments of $y_{ni}^2$. To this end,  
\begin{eqnarray}\nonumber
\E y_{ni}^2 &=&n^{-1}\E~\E\left( \sum_{1\le j,k\le n}
\left(\eta_{ij}-\mu(x_i)\right)\left(\eta_{ik}-\mu(x_i)\right)|x_i\right)
=
\E~\E \left( \sum_{1\le j\le n} \left(\eta_{ij}-\mu(x_i)\right)^2|x_i\right)\\
\lbl{stop-here}
&+&
2n^{-1} \E~\E\left( \sum_{1\le j<k\le n} \left(\eta_{ij}-\mu(x_i)\right)
\left(\eta_{ik}-\mu(x_i)\right)|x_i\right).
\end{eqnarray}
The first term on the right hand side of (\ref{stop-here}) is equal to $\sigma^2$
(see (\ref{Veta})). The second term is equal to $0$, as easy to see using the
independence
of $\eta_{ij}-\mu(x_i)$ and $\eta_{ik}-\mu(x_i)$ for $k\neq j$.
Thus,
\be\lbl{rhs-Ey2}
\E y_{ni}^2=\sigma^2+
2n^{-1} \E\left(  \sum_{1\le j<k\le n}
\E\left(\eta_{ij}-\mu(x_i)|x_i\right)\E\left(\eta_{ik}-\mu(x_i)| x_i\right)\right)=\sigma^2.
\ee
Recall that $\sigma^2>0$, by (\ref{1-positive-sigma}).
Similarly, we compute 
\begin{eqnarray}\nonumber
\E(y_{ni}^4)&=&n^{-2}\E~\E\left( \sum_{1\le j_1,j_2,j_3,j_4\le n}
\left(\eta_{i{j_1}}-\mu(x_i)\right)\dots\left(\eta_{ij_4}-\mu(x_i)\right)|x_i\right)\\
&=&
\nonumber
6n^{-2}\E\left(  \sum_{1\le j<k\le n}  \E\left(\eta_{ij}-\mu(x_i)| x_i\right)^2
E \left(\eta_{ik}-\mu(x_i)|x_i\right)^2\right)+
n^{-2}\E\left(  \sum_{1\le j\le n} \E\left(\eta_{ij}-\mu(x_i)|x_i\right)^4\right)\\
\lbl{rhs-Ey4}
&=& {6n(n-1)\over n^2}   \sigma^4+O(n^{-1})=6\sigma^4+O(n^{-1})
\end{eqnarray}
and 
\begin{eqnarray}\nonumber
\E y_{ni}^6&=&n^{-3}\E ~\E\left( \sum_{1\le j_1,j_2,j_3,j_4,j_5,j_6\le n}
\left(\eta_{i{j_1}}-\mu(x_i)\right)\dots\left(\eta_{ij_6}-\mu(x_i)\right)|x_i\right)\\
&=&
\nonumber
{ 6 \choose 2} {4\choose 2} n^{-3}\E\left(  \sum_{1\le j<k<l\le n}  \E\left(\eta_{ij}-\mu(x_i)|x_i\right)^2
E \left(\eta_{ik}-\mu(x_i)|x_i\right)^2    E \left(\eta_{il}-\mu(x_i)|x_i\right)^2 \right)+O(n^{-1})\\
\lbl{rhs-Ey6}
&=& {90n(n-1)(n-2)\over n^3} \sigma^6+O(n^{-1})=90\sigma^6+O(n^{-1}).
\end{eqnarray}

For $n\in\N$, let
\be\lbl{define-zeta}
\zeta_{ni}:={ y^2_{ni} - \E y_{ni}^2\over  \sqrt{n \V(y^2_{in} )}}=
{ y_{ni}^2-\sigma^2 \over   n^{1/2}\sqrt{ 5\sigma^4+O(n^{-1})}   },\; i\in [n],
\ee
where (\ref{rhs-Ey2}) and (\ref{rhs-Ey4}) were used to obtain the expression
on the right hand side.

Consider
\be\lbl{seq-zeta}
\zeta_{n1}, \zeta_{n2},\dots,\zeta_{nn}.
\ee
By construction, $\zeta_{ni}, i\in [n]$, are IID RVs. Further,
\be\lbl{centered}
\E\zeta_{ni}=0 \;\mbox{and}\; \V\left(\sum_{i=1}^{n}\zeta_{ni}\right)=1.
\ee
Moreover, the triangular array (\ref{seq-zeta}) satisfies the Lyapunov condition \cite{Bill-Prob}
\be\lbl{Lyap}
\sum_{i=1}^n \E |\zeta_{ni}|^3\le 
{ \sum_{i=1}^n  \E\left( y_{ni}^6+3y_{ni}^4\sigma^2+3y_{ni}^2\sigma^4+\sigma^6\right)
\over n^{3/2} \left(5\sigma^4+O(n^{-1})\right)^{3/2}}=O(n^{-1/2})\to 0\;
\mbox{as}\; n\to \infty.
\ee

From (\ref{centered}) and (\ref{Lyap}), via the CLT,
we conclude that
\be\lbl{apply-clt}
{ \sum_{i=1}^n (y_{ni}^2-\sigma^2) \over  \sqrt{ n( 5\sigma^4+O(n^{-1}))}   }
={ n^{-1}\sum_{i=1}^n y_{ni}^2-\sigma^2 \over  
n^{-1/2}\sqrt{  5\sigma^4+O(n^{-1})}   }
\overset{d}{\longrightarrow} \mathcal{N}(0,1)\; n\to\infty.
\ee
The statement  (\ref{clt}) follows from (\ref{apply-clt}) and 
the definition of $y_{ni}$ (\ref{def-yni}).\\
$\qed$

For the proof of Theorem~\ref{thm.W-convergence}, we need to extend
Lemma~\ref{lem.clt} to cover the case when $f$ depends on $t\in [0,T]$
in addition to $(x,y)\in I^2$.

\begin{cor}\lbl{extend-lem.clt}
Suppose that $f$ in Lemma~\ref{lem.clt} also depends on $t\in [0,T]$,
and $\mathbf{f}\in C(0,T; L^\infty(I^2))$ if viewed as a mapping 
from $[0,T]$ to $L^\infty(I^2)$, $\mathbf{f}(t)=f(\cdot,t)\in
L^\infty(I^2)$. Adding $t-$dependence to all variables  defined
using $f$ and, otherwise, keeping the notation of Lemma~\ref{lem.clt},
we assume that
\be\lbl{assume-extend}
  \min_{t\in [0,T]} \sigma^2(t)\ge c_1>0.
\ee 
Then the conclusion of Lemma~\ref{lem.clt} holds for $t-$dependent sums 
for every $t\in [0,T]$
\be\lbl{now-cor}
{ S_n(t)-\sigma_n^2(t) \over  n^{-1/2}\sqrt{ 5\sigma_n^4(t)+O(n^{-1})}   }
\overset{d}{\longrightarrow} \mathcal{N}(0,1)\; n\to\infty.
\ee
\end{cor}
\pf\;
From the assumption $\mathbf{f}\in C(0,T; L^\infty(I^2))$ and (\ref{assume-extend}),
for 
$$
\sigma^2(t)=\int_{I^2} f(x,y,t)^2 W(x,y)dxdy -\int_I\left(\int_I f(x,y,t)W(x,y) dy\right)^2 dx,
$$
we have 
\be\lbl{need-this-bound}
0<c_1\le \sigma^2(t)\le 2\|\mathbf{f}\|^2_{C(0,T; L^\infty(I^2))}.
\ee
With these bounds, by repeating the steps in the proof of Lemma~\ref{lem.clt}, we first
show that  $t-$dependent moments of $y_{ni}^2(t)$  are bounded uniformly in $t\in [0,T]$;
then verify Lyapunov condition for every $t\in [0,T]$ and apply the CLT. This shows 
(\ref{now-cor}).
$\qed$

We are now in a position to prove Theorem~\ref{thm.W-convergence}.\\
\pf (Theorem~\ref{thm.W-convergence})
Denote $\zeta_{ni}(t)= u(x_i,t) -u_{ni}(t),\; i\in [n]$ and let
$$
\zeta_n(t)=\left(\zeta_{n1}(t), \zeta_{n2}(t),\dots, \zeta_{nn}(t)\right).
$$
By subtracting Equation~$i$ in (\ref{Wheat}) from the corresponding
equation in (\ref{Wcont}) evaluated  at $x=x_i$, we have
\be\lbl{derive-xi}
{d\over dt} \zeta_{ni}(t)= z_{ni}(t)
+ 
{1\over n} \sum_{j=1}^n \xi_{ij} \left[ D\left(u(x_j,t)-u(x_i,t)\right) - 
D\left(u_{nj}(t)-u_{ni}(t)\right)\right],
\ee
where
\be\lbl{where-zni}
z_{ni}=
\int_I  W(x_i,y) D\left(u(y,t)-u(x_i,t)\right)dy- 
{1\over n} \sum_{j=1}^n \xi_{ij} D\left(u(x_j,t)-u(x_i,t)\right),
\ee
and $\xi_{ij}$ are defined in (\ref{define-xiij}).

Next, we multiply both sides of (\ref{derive-xi}) by $n^{-1}\zeta_{ni}$ and sum
over $i$ to obtain
\be\lbl{pre-Gron-1}
{1\over 2}{d\over dt} \|\zeta_n\|_{2,n}^2= (z_n, \zeta_n)_{n} + {1\over n^2} \sum_{i,j=1}^n
\xi_{ij} \left[ D\left(u(x_j,t)-u(x_i,t)\right) - 
D\left(u_{nj}(t)-u_{ni}(t)\right)\right]\zeta_{ni},
\ee
where $z_n=(z_{n1}, z_{n2},\dots,z_{nn})$.
We estimate the first term on the right-hand side of (\ref{pre-Gron-1}) via the Cauchy-Schwarz
inequality
\be\lbl{use-CS}
\left|(z_n, \zeta_n)_{n} \right|\le \|z_n\|_{2,n} \|\zeta_n\|_{2,n}\le 2^{-1}(\|z_n\|_{2,n}^2 +
\|\zeta_n\|_{2,n}^2).
\ee
For the second term we  use the Lipschitz continuity of $D$,
$|\xi_{ij}|\le 1$, the Cauchy-Schwarz inequality, and the triangle inequality to obtain
$$
\left|
{1\over n^2} \sum_{i,j=1}^n
\xi_{ij} \left[ D\left(u(x_j,t)-u(x_i,t)\right) - 
D\left(u_{nj}(t)-u_{ni}(t)\right)\right]\zeta_{ni}
\right|\le 
$$
\be\lbl{pre-Gron-2}
{L\over n^2} \sum_{i,j=1}^n \left(|\zeta_{nj}(t)|+|\zeta_{ni}(t)|\right) |\zeta_{ni}(t)|
\le 2L \|\zeta_n(t)\|^2_{2,n}.
\ee
Using (\ref{pre-Gron-1}), (\ref{use-CS}), and (\ref{pre-Gron-2}), we have
\be\lbl{pre-Gron-3}
{d\over dt} \|\zeta_n\|^2_{2,n} \le (4L+1) \|\zeta_n\|^2_{2,n} +\|z_n\|_{n,2}^2.
\ee 
From (\ref{pre-Gron-3}) via the Gronwall's inequality we have
\be\lbl{apply-Gronwall}
\sup_{t\in[0,T]} \|\zeta_n(t)\|_{2,n} \le {\sup_{t\in[0,T]} \|z_n(t)\|^2_{2,n}\over
4L+1} \exp\{ (4L+1)T\}.
\ee

It remains to bound $\sup_{t\in[0,T]} \|z_n(t)\|^2_{2,n}.$
To this end, let
$$
f(x,y,t):=D(u(y,t)-u(x,t)).
$$
Using $\mathbf{u}\in C(0,T; L^\infty(I))$, Lipschitz continuity of $D$, and
the triangle inequality, we have
\be\lbl{uniform}
\|\mathbf{f}\|_{C(0,T;L^\infty(I^2))}\le L \max_{t\in [0,T]} 
\esssup_{(x,y)\in I^2} \left| u(x,t)-u(y,t)\right|\le
2L \|\mathbf{u}\|_{C(0,T; L^\infty (I))}.
\ee
By (\ref{new-assume}) and (\ref{uniform}), we find that $\sigma^2(t)$ is bounded
for $t \in[0,T]$
\be\lbl{find-bdd}
C_1\le \sigma^2(t)\le 2L\|\mathbf{u}\|_{C(0,T; L^\infty (I))}=:C_2.
\ee

Using Corollary~\ref{extend-lem.clt}, for $z_n=(z_{n1}, z_{n2},\dots,z_{nn})$ (see (\ref{where-zni})),
we have
$$
{n\|z_n\|_{2,n (t)}^2 -\sigma^2(t)\over n^{-1/2}\beta(\sigma^2(t))} 
\stackrel{d}{\rightarrow} \mathcal{N} (0,1),\;\mbox{where}\; 
\beta(\sigma^2(t))=\sqrt{ 5\sigma^2(t)+O(n^{-1})}.
$$
Further, we have 
\begin{eqnarray}\nonumber
\P\left( \left| n \|z_n(t)\|_{2,n}^2 -\sigma^2(t)\right|>1\right)& =&
\P\left( {\left| n \|z_n(t)\|_{2,n}^2 -\sigma^2(t)\right|\over n^{-1/2} \beta(\sigma^2(t))}>
{n^{1/2}\over \beta(\sigma^2(t))}\right)\\
\lbl{ap-clt}
&\le&
\P\left( {\left| n \|z_n(t)\|_{2,n}^2 -\sigma^2(t)\right|\over n^{-1/2} \beta(\sigma^2(t))}>
{n^{1/2}\over C_2}\right)
\rightarrow 0,
\end{eqnarray}
as $n\to\infty$. We used (\ref{find-bdd}) to obtain the last
inequality in (\ref{ap-clt}).  Convergence in (\ref{ap-clt}) is uniform for $t\in [0,T]$.
Therefore, $\|z_n(t)\|^2_{n,2}$ converges to zero in probability uniformly in $t$.
Moreover,
$$
\P\left( \|z_n(t)\|^2_{n,2}> (C_2+1) n^{-1} \right)\le
\P\left( \left| n \|z_n(t)\|_{2,n}^2 -\sigma^2(t)\right|>1\right)\rightarrow 0,
\;\mbox{as}\; n\to\infty
$$
uniformly for $t\in [0,T]$.

Let $\epsilon>0$ be arbitrary. Then for $C_3:=C_2+1$ and for some $N\in\N$,
we have 
$$
\P\left(\sup_{t\in [0,T]}\|z_n(t)\|^2_{2,n}>C_3n^{-1}\right) <\epsilon \;\mbox{for}\; n>N.
$$  
The combination of this and (\ref{apply-Gronwall}) proves the theorem.\\
$\qed$

\section{Networks on W-random graphs generated by deterministic sequences}
\lbl{sec.det}
\setcounter{equation}{0}

In this section, we consider the heat equations on $W-$random graphs 
generated by deterministic sequences of points from $I$.
To this end, we partition $I$ into $n$ subintervals
\be\lbl{partition}
I_{ni}=[(i-1)n^{-1}, in^{-1}), \; i\in [n-1],\; \mbox{and}\;I_{nn}=[(n-1)n^{-1},
1].
\ee
Suppose 
\be\lbl{set-Xn}
X_n=\{x_{n1}, x_{n2},\dots, x_{nn}\},\; x_{ni}\in \bar I_{ni} \; i\in [n],
\ee
where $\bar I_{ni}$ denotes the closure of $I_{ni}$.

\begin{df}\lbl{df.Wdet}
Graph $G_n=\langle V(G_n), E(G_n)\rangle$ is called a $W$-random graph 
generated by the deterministic sequence $X_n$ and is  denoted 
$G_n=\mathbb{G}(W,X_n)$, if $V(G_n)=[n]$ and
for every $(i,j)\in [n]^2,\; i\neq j,$
$$
\P\left\{ (i,j)\in E(G_n)\right\} =W(x_i, x_j).
$$
The decision whether to include $(i,j)$ to $E(G_n)$  is made
independently for each pair $(i,j)\in [n]^2,\; i\neq j$.
\end{df}

\begin{rem} If $W$ is continuous on $I$ almost everywhere, then 
$\{\mathbb{G}(W,X_n)\}$ is convergent with the limit given by graphon $W$ 
(cf.~Lemma~2.5 \cite{BorChay11}).
\end{rem}

Let  $u_n(t)=(u_{n1}(t), u_{n2}(t),\dots,u_{nn}(t))$ denote
the solution of the IVP (\ref{Wheat}),
(\ref{Wheat-ic}) for the heat equation on $G_n=\mathbb{G}(W,X_n)$, 
and  define $u_n: I\times \R\to \R$ as follows. For $x\in I_{ni},$
$i\in [n],$ let
$$
u_n(x,t)=u_{ni}(t),  t\in\R.
$$

\begin{thm}\lbl{thm.Wdet}
Suppose $W\in\mathcal{W}_0$ is almost everywhere continuous on $I^2$, $D:~\R\to\R$
is Lipschitz continuous, and $g\in L^\infty(I)$.
Let $u(x,t)$ denote the solution of the IVP (\ref{Wcont}), (\ref{Wcont-ic}).
Suppose further
\footnote{Because $\mathbf{u}\in C(\R,L^\infty(I))$, $D$
  is Lipschitz, and $W$ is bounded, the integral in
  (\ref{positive-sigma}) defines a continuous function of $t$. Thus, 
the use $\min$ in (\ref{positive-sigma}) is justified.}
\be\lbl{positive-sigma}
\min_{t\in [0,T]} \int_{I^2} D(u(y,t)-u(x,t)) W(x,y)(1-W(x,y))dxdy >0.
\ee
Then for any $T>0$
\be\lbl{L-estimate}
\left\|\mathbf{u}_n-\mathbf{u}\right\|_{C(0,T;L^2(I))}\mathop{\rightarrow}^{p} 0\;
\mbox{as}\; n\to\infty.
\ee
The convergence in (\ref{L-estimate}) is in probability.
\end{thm}

For the proof of Theorem~\ref{thm.Wdet} we need to derive several
auxiliary results.  The first result is parallel to Lemma~\ref{lem.clt} of the previous section. 

\begin{lem}\lbl{2-lem.clt}
Let $\{ W_{nij}\}$ and $\{f_{nij}\}$ be two real arrays defined for $n\in\N$ 
and $i,j \in [n]$, and
\begin{eqnarray}\lbl{def-sigma-ni}
\sigma_{ni}^2& =& n^{-1}\sum_{i=1}^n f_{nij}^2 W_{nij}(1-W_{nij}), i\in [n],\\  
\lbl{def-sigma-n}
\sigma_n^2&=&n^{-1}\sum_{i=1}^n \sigma_{ni}^2.
\end{eqnarray}
Assume   that $\{f_{nij}\}$, $n\in\N, i,j\in [n]$, is a bounded array, $0\le w_{nij}\le 1$ and
\be \lbl{2-positive-sigma}
\liminf_{n\to\infty}\sigma_n^2>0.
\ee

Let $\{\xi_{nij}\}, n\in\N, (i,j)\in [n]^2$ be independent binomial RVs
$\mathcal{L}\left(\xi_{nij} \right) = \mbox{Bin}( W_{nij})$.
Further, let
\begin{eqnarray*}
\eta_{nij}& =&\xi_{nij} f_{nij}, \; (i,j)\in [n]^2,\\
z_{ni}&=&{1\over n}\sum_{j=1}^n \left(\eta_{nij} - f_{nij}W_{nij}\right), \\
S_n &=&\sum_{i=1}^n z_{ni}^2.
\end{eqnarray*}

Then 
\be\lbl{2-clt}
{ S_n-\sigma_n^2 \over  n^{-1/2}\sqrt{ 5\sigma_n^4+O(n^{-1})}   }
\overset{d}{\longrightarrow} \mathcal{N}(0,1) \;\mbox{as}\; n\to\infty.
\ee
\end{lem}

\pf 
First, compute the moments of  the independent RVs $\{\eta_{nij}\}$, $n\in\N$, $(i,j)\in [n]^2$, 
$$
\E\eta_{nij}^k=f^k_{nij}W_{nij}, \; k\in\N.
$$
Thus, for $y_{ni}=\sqrt{n}z_{ni},$  $i\in [n]$, we have $\E y_{ni} = 0$. 
Further,
\begin{eqnarray*}
\E y_{ni}^2&=&n^{-1} \E\left( \sum_{1\le j,k\le n}
\left(\eta_{nij}-f_{nij}W_{nij} \right)\left(\eta_{nik}-f_{nij}W_{nij}\right)\right) \\
&=&  n^{-1}\E \left( \sum_{1\le j\le n} \left(\eta_{nij}- f_{nij}W_{nij}\right)^2\right)+
2n^{-1} \E\left( \sum_{1\le j<k\le n} \left(\eta_{nij}-   f_{nij}W_{nij}\right)
\left(\eta_{ik}-f_{nik}W_{nik}\right)\right)\\
&=&\sigma_{ni}^2+
2n^{-1} \sum_{1\le j<k\le n}
\E\left(\eta_{nij}- f_{nij}W_{nij}\right)\E\left(\eta_{nik}-f_{nik}W_{nik}\right)=\sigma_{ni}^2,
\end{eqnarray*}
where
\be\lbl{sigmani}
\sigma_{ni}^2:=n^{-1}\E \left( \sum_{1\le j\le n} \left(\eta_{nij}- f_{nij}W_{nij}\right)^2\right)
=n^{-1} \sum_{j=1}^n f_{nij}^2 W_{nij} \left(1-W_{nij}\right).
\ee

Similarly, we compute 
\begin{eqnarray*}
\E y_{ni}^4 &=&n^{-2} \E\left( \sum_{1\le j_1,j_2,j_3,j_4\le n}
\left(\eta_{ni{j_1}}-f_{nij_1}W_{nij_1} \right)\dots\left(\eta_{nij_4}- 
f_{nij_4}W_{nij_4}\right)\right)\\
&=&
6n^{-2} \sum_{1\le j<k\le n}  \E\left(\eta_{nij}-f_{nij}W_{nij}\right)^2
E \left(\eta_{nik}- f_{nik}W_{nik}\right)^2+
n^{-2}  \sum_{1\le j\le n} \E\left(\eta_{nij}-f_{nij}W_{nij}\right)^4\\
&=& {6n(n-1)\over n^2}   \sigma_{ni}^4+O(n^{-1})=6\sigma_{ni}^4+O(n^{-1}).
\end{eqnarray*}
and 
\begin{eqnarray*}
\E y_{ni}^6 &=&n^{-3}\E \left(   \sum_{1\le j_1,j_2,j_3,j_4,j_5,j_6\le n}
\left(\eta_{ni{j_1}}-f_{nij_1}W_{nij_1}\right)\dots\left(\eta_{nij_6}-f_{nij_6}W_{nij_6}\right)\right)\\
&=&
{ 6 \choose 2} {4\choose 2} n^{-2}\sum_{1\le j<k<l\le n}  
\E\left(\eta_{nij}- f_{nij}W_{nij}\right)^2
E \left(\eta_{nik}- f_{nik}W_{nik}\right)^2  
\E \left(\eta_{nil}- f_{nil}W_{nil}\right)^2 +O(n^{-1})\\
&=& {90n(n-1)(n-2)\over n^3} \sigma_{ni}^6+O(n^{-1})=90\sigma_{ni}^6+O(n^{-1}).
\end{eqnarray*}

For $n\in\N$, let
\be\lbl{define-zeta}
\zeta_{ni}:={ y^2_{ni} - \E y_{ni}^2\over  \sqrt{n \V(y^2_{in} )}}=
{ y_{ni}^2-\sigma_{ni}^2 \over   n^{1/2}\sqrt{ 5\sigma_{ni}^4+O(n^{-1})}   },\; i\in [n].
\ee

Consider
\be\lbl{2-seq-zeta}
\zeta_{n1}, \zeta_{n2},\dots,\zeta_{nn}.
\ee
By construction, $\zeta_{ni}, i\in [n]$ are independent RVs. Further,
\be\lbl{2-centered}
\E\zeta_{ni}=0 \;\mbox{and}\; \V\left(\sum_{i=1}^{n}\zeta_{ni}\right)=1.
\ee
Moreover, the triangular array (\ref{seq-zeta}) satisfies the Lyapunov condition \cite{Bill-Prob}
\be\lbl{2-Lyap}
\sum_{i=1}^n \E |\zeta_{ni}|^3\le 
{ \sum_{i=1}^n  \E\left( y_{ni}^6+3y_{ni}^4\sigma_{ni}^2+
3y_{ni}^2\sigma_{ni}^4+\sigma_{ni}^6\right)
\over n^{3/2} \left(5\sigma_n^4+O(n^{-1})\right)^{3/2}}=O(n^{-1/2})\to 0\;
\mbox{as}\; n\to \infty.
\ee

From (\ref{2-centered}) and (\ref{2-Lyap}), using the CLT,
we conclude that
\be\lbl{2-apply-clt}
{ \sum_{i=1}^n (y_{ni}^2-\sigma_{ni}^2) \over  
\sqrt{ n( 5\sigma_n^4+O(n^{-1}))}   }
={ n^{-1}\sum_{i=1}^n y_{ni}^2-\sigma_n^2 \over  
n^{-1/2}\sqrt{  5\sigma_n^4+O(n^{-1})}   }
\overset{d}{\longrightarrow} \mathcal{N}(0,1),\; n\to\infty.
\ee
The statement  (\ref{2-clt}) follows from (\ref{2-apply-clt}) and 
the definition of $y_{ni}$.\\
$\qed$

With obvious modifications the proof of Lemma~\ref{2-lem.clt} can be easily extended
to cover the following version of the lemma.
\begin{cor}\lbl{t-dependence}
Suppose $f_{nij}$ in Lemma~\ref{2-lem.clt} depend on real parameter $t\in [0,T]$
for some $T$.
Keeping the notation of Lemma~\ref{2-lem.clt}, we add $t-$dependence
to all variables defined using $f_{nij}(t)$. Assume that functions
$f_{nij}(t)$, $n\in\N, i,j,\in [n]$, are uniformly bounded for $t\in [0,T]$ and 
\be\lbl{bound-from-below}
\liminf_{n\to\infty}\sigma_n^2(t)=\liminf_{n\to\infty} n^{-1}
\sum_{i,j=1}^n f_{nij}(t)W_{nij}(1-W_{nij})
\ge C_1 >0
\ee
for every $t\in [0,T]$.

Then the conclusion of Lemma~\ref{2-lem.clt} holds for $t-$dependent sums for every $t\in [0,T]$
$$
{ S_n(t)-\sigma_n^2(t) \over  n^{-1/2}\sqrt{ 5\sigma_n^4(t)+O(n^{-1})}   }
\overset{d}{\longrightarrow} \mathcal{N}(0,1), \; n\to\infty.
$$
\end{cor}

Having prepared the application of the CLT that will be needed in the proof of Theorem~\ref{thm.Wdet},
we now introduce an auxiliary IVP for the heat equation on a weighted graph $\tilde G_n=\mathbb{H}(W, X_n)$.
The latter is a complete graph on $n$ nodes, $V(\tilde G_n)=[n]$. Each edge of $\tilde G_n$ is supplied 
with the weight
$$
W_{nij}=W(x_{ni},x_{nj}), \; (i,j)\in [n]^2,\; i\neq j.
$$

Consider the IVP for the heat equation on the weighted graph $\tilde G_n$ 
\begin{eqnarray}\lbl{weight}
{d\over dt} v_{ni}(t) &=& n^{-1} \sum_{j: (i,j)\in E(\tilde G_n)} W_{nij} D(v_{nj}-v_{ni}),\\
\lbl{weight-ic}
v_{ni}(0)&=&g(x_i),\; i\in [n].
\end{eqnarray}
Denote the solution of the IVP (\ref{weight}) and (\ref{weight-ic}) by 
$v_n(t)=(v_{n1}(t), v_{n2}(t),\dots,v_{nn}(t))$.
Let $v_n(x,t)$ be a function defined on $I\times \R$ and such that for
$x\in I_{ni}, i\in [n]$
$$
v_n(x,t)=v_n(t), \; t\in\R.
$$
Next, define a step-function $W_n$ on $I^2$ such that for $(x,y)\in I_{ni}\times I_{nj}$,
$i,j\in [n],$ 
$$
W_n(x,y)=W_{nij}.
$$
By construction, $v_n(x,t)$ solves the following IVP
\begin{eqnarray}\lbl{step-IVP}
{\p \over \p t} v_n(x,t) &=& \int_I W_n(x,y) D(v_n(y,t)-v_n(x,t))  dy,\\
\lbl{step-IVP-ic}
v_n(x,0) &=& g(x_{ni}),\; x\in I_{ni}, i\in [n].
\end{eqnarray}
It was shown in \cite{GKVM13} that for large $n$,  $v_n(x,t)$ approximates
the solution of the IVP (\ref{Wcont}), (\ref{Wcont-ic}). Specifically, we have the 
following lemma. 

\begin{lem}\cite[Theorem 5.2]{GKVM13}\lbl{lem.recall}
Suppose $W\in L^\infty(I^2)$ is almost everywhere continuous on $I^2$,
$D$ is Lipschitz continuous, 
and $g\in L^\infty(I)$. 
Then for any $T>0$
\be\lbl{recall}
\| \mathbf{u} - \mathbf{v}_n\|_{C(0,T; L^2(I))}\to 0 \;\mbox{as}\; n\to \infty.
\ee
\end{lem}

We use Lemma~\ref{lem.recall} to derive the following result.

\begin{lem}\lbl{sigma-n-converge}
Suppose $W\in\mathcal{W}_0$ is almost everywhere continuous on $I^2$,
$D$ is Lipschitz continuous, 
and $g\in L^\infty(I)$. 
Let $u(x,t)$ and $v_n(x,t)$ denote the solutions of the IVPs (\ref{Wcont}), (\ref{Wcont-ic})
and (\ref{step-IVP}), (\ref{step-IVP-ic}), respectively; and let  
\begin{eqnarray*}
\sigma^2(t)&=&\int_{I^2} D(u(y,t)-u(x,t)) W(x,y)(1-W(x,y))dxdy,\\
\sigma_n^2(t)&=& \int_{I^2} D(v_n(y,t)-v_n(x,t)) W_n(x,y)(1-W_n(x,y))dxdy.
\end{eqnarray*}
Then
$$
\sup_{t\in [0,T]} \left|\sigma^2_n(t)-\sigma^2(t)\right|\le 
C_2\left[ \|\mathbf{v}_n-\mathbf{u}\|_{C(0,T;L^2(I))}+\|W_n-W\|_{L^2(I^2)}\right],
$$
for some $C_2>0$. In particular, $\sigma_n^2\to \sigma^2$ uniformly for $t\in [0,T]$.
\end{lem}

\pf\;
\begin{enumerate}
\item Using Lipschitz continuity of $D$ and the triangle inequality, for any $t\in [0,T]$ we have
\begin{eqnarray}\nonumber
\left|\int_{I^2} D(v_n(y,t)-v_n(x,t))-D(u(y,t)-u(x,t))dxdy\right| 
\le L \int_{I^2}  |v_n(y,t)-u(y,t)| \\
\lbl{step-1}
+|v_n(x,t)-u(x,t)|  dxdy\le 
2L\|\mathbf{v}_n-\mathbf{u}\|_{C(0,T;L^2(I))}\to 0,
\end{eqnarray}
as $n\to\infty$. Therefore,
\be\lbl{step-2}
\max_{t\in[0,T]}\left|\int_{I^2} D(v_n(y,t)-v_n(x,t)) dxdy\right| \le C_3,\; n\in\N,
\ee
for some $C_3$ independent of $n$.
\item Denote $q(x)=x(1-x)$. For $x,y\in [0,1]$, $|q(x)-q(y)|\le |x-y|$. Thus,
\be\lbl{step-3}
|q(W)-q(W_n)|\le |W-W_n|.
\ee
\item Finally, we estimate $\left|\sigma_n(t)-\sigma (t)\right|$. For arbitrary $t\in [0,T]$,
we have
\begin{eqnarray}
\nonumber
&&\left|\int_{I^2} D(v_n(y,t)-v_n(x,t)) q(W_n(x,y))dxdy  -
\int_{I^2} D(u(y,t)-u(x,t)) q(W(x,y))dxdy \right| \\
\nonumber
&\le& 
\left|\int_{I^2} D(v_n(y,t)-v_n(x,t)) \left[ q(W_n(x,y))-q(W(x,y))\right] dxdy\right|\\
\lbl{step-4}
&+&\left|\int_{I^2}\left[ D(v_n(y,t)-v_n(x,t))- D(u(y,t)-u(x,t))\right] q(W(x,y))dxdy \right|. 
\end{eqnarray}
Using the Cauchy-Schwarz inequality, Lipschitz continuity of $D$, $|q(W)|\le 1$,
 (\ref{step-2}), and (\ref{step-3}) from (\ref{step-4}) we  obtain
\be\lbl{delta-sigma}
\sup_{t\in [0,T]}\left|\sigma_n(t)-\sigma (t)\right|\le C_3 \|W-W_n\|_{L^2(I^2)} + 
L \|\mathbf{v}_n-\mathbf{u}\|_{C(0,T;L^2(I))}.
\ee

Note that $W_n\to W$ as $n\to\infty$ at every point of continuity of $W$, i.e., almost everywhere
on $I^2$. Therefore, by the dominated convergence theorem, 
\be\lbl{Wn-converge}
\|W-W_n\|_{L^2(I^2)} \to 0\; \mbox{as} \; n\to \infty.
\ee
The statement of the lemma follows from (\ref{delta-sigma}), (\ref{Wn-converge}), and 
Lemma~\ref{lem.recall}.\; $\qed$
\end{enumerate}

\pf (Theorem~\ref{thm.Wdet})
Denote $\eta_{ni}(t)= u_{ni}(t) -v_{ni}(t),\; i\in [n],$ and
$$
\eta_n(t)=\left(\eta_{n1}(t), \eta_{n2}(t),\dots, \eta_{nn}(t)\right).
$$
By subtracting Equation~$i$ in (\ref{weight}) from the corresponding
equation in (\ref{Wheat}) written for $G_n=\mathbb{G}(W,X_n)$, we have
\be\lbl{derive-eta}
{d\over dt} \eta_{ni}= 
{1\over n} \left( \sum_{j=1}^n \xi_{nij}  D(u_{nj}-u_{ni}) - \sum_{j=1}^n W_{nij}  D(v_{nj}-v_{ni})\right).
\ee
By rewriting the right-hand side of (\ref{derive-eta}), we obtain
\be\lbl{rewrite-eta}
 {d\over dt} \eta_{ni}= 
{1\over n} \sum_{j=1}^n \xi_{nij} \left[ D(u_{nj}-u_{ni}) - D(v_{nj}-v_{ni})\right]+z_{ni},
\ee
where
\be\lbl{def-zni}
z_{ni}={1\over n} \sum_{j=1}^n \xi_{nij}  D(v_{nj}-v_{ni}) - {1\over n} \sum_{j=1}^n w_{nij}  D(v_{nj}-v_{ni}).
\ee

By multiplying both sides of (\ref{rewrite-eta}) by $n^{-1}\eta_{ni}$ and summing over $i$, we have
\be\lbl{next-eta}
{1\over 2}{d\over dt} \|\eta_n\|_{2,n}^2= {1\over n^2} \sum_{i,j=1}^n
\xi_{ij} \left[ D(u_{nj}-u_{ni}) - D(v_{nj}-v_{ni})\right]\eta_{ni} + (z_n, \eta_n)_{n}.
\ee
We bound the first term on the right hand side of (\ref{next-eta})  using the Lipschitz continuity of $D$,
$|\xi_{ij}|\le 1$, the Cauchy-Schwarz inequality, and the triangle inequality 
$$
\left|
{1\over n^2} \sum_{i,j=1}^n
\xi_{ij} 
\left[ D(u_{nj}-u_{ni}) - D(v_{nj}-v_{ni})\right]\eta_{ni} 
\right|\le 
$$
\be\lbl{eta-preGron}
{L\over n^2} \sum_{i,j=1}^n \left(|\eta_{nj}|+|\eta_{ni}|\right) |\eta_{ni}|
\le 2L \|\eta_{ni}\|^2_{2,n}.
\ee
We bound the  second term using the Cauchy-Schwarz inequality
\be\lbl{bound-second}
\left|(z_n, \eta_n)_{n} \right|\le \|z_n\|_{2,n} \|\eta_n\|_{2,n}\le {1\over 2} 
\left(\|z_n\|_{2,n}^2+ \|\eta_n\|_{2,n}^2\right),
\ee
where  $z_n=(z_{n1}, z_{n2},\dots, z_{nn}).$

The combination of (\ref{next-eta}), (\ref{eta-preGron}), and (\ref{bound-second}) yields
 \be\lbl{Gron-ready}
{d\over dt} \|\eta_n\|^2_{2,n} \le (4L+1) \|\eta_n\|^2_{2,n} + \|z_n\|^2_{2,n}.
\ee
By Gronwall's inequality,
\be\lbl{use-Gron}
\max_{t\in[0,T]} \|\eta\|^2_{2,n} \le {\max_{t\in[0,T]} \|z_n(t)\|^2_{2,n}\over 4L+1} \exp\{ (4L+1)T\}.
\ee
Thus,
\be\lbl{use-Gron-1}
\max_{t\in[0,T]} \|\eta\|_{2,n} \le {\max_{t\in[0,T]} \|z_n(t)\|_{2,n}\over \sqrt{4L+1}} \exp\{ (2L+1)T\}.
\ee

It remains to estimate $\|z_n(t)\|_{2,n}$ (see (\ref{def-zni})). To this end, we use 
Corollary~\ref{t-dependence} with
$$
f_{nij}(t)= D(v_{nj}(t)-v_{ni}(t)) \; \mbox{and}\; W_{nij}=W(x_{ni}, x_{nj}).
$$
From  Lemma~\ref{sigma-n-converge} and (\ref{positive-sigma}), we have
\be\lbl{bound-sigma-n-below}
\min_{t\in [0,T]} \sigma_n^2(t)\ge C_4>0,
\ee
for sufficiently large $n$.
In particular, (\ref{bound-from-below}) holds.
Similarly, by  Lemma~\ref{sigma-n-converge}, we have 
\be\lbl{bound-sigma-n}
\max_{t\in [0,T]} \sigma_n^2(t)\le C_5, \; n\in \N.
\ee

By Corollary~\ref{t-dependence}, for arbitrary $t\in [0,T]$, we have
$$
\P\{ |n\|z_n(t)\|_{2,n}^2-\sigma_n^2(t)|>1\} = 
\P\left\{ \left|{ n\|z_n(t)\|_{2,n}^2-\sigma_n^2(t)\over
n^{-1/2} \sqrt{5\sigma_n^4(t)+O(n^{-1})}}\right|>
{n^{1/2}\over \sqrt{5\sigma_n^4(t)+O(n^{-1})}}\right\}
$$
\be\lbl{pre-bound-zn}
\le\P\left\{ \left|{ n\|z_n(t)\|_{2,n}^2-\sigma_n^2(t)\over
n^{-1/2} \sqrt{5\sigma_n^4(t)+O(n^{-1})}}\right|>
{n^{1/2}\over \sqrt{5 C_5^2+O(n^{-1})}}\right\}\to 0 \;\mbox{as}\; n\to\infty.
\ee

Using (\ref{bound-sigma-n}), from (\ref{pre-bound-zn}) we have
\be\lbl{bound-zn}
\P\{ \|z_n(t)\|_{2,n}^2\le (C_5+1)n^{-1}\} \le \P\{ |n\|z_n(t)\|_{2,n}^2-\sigma_n^2(t)|>1\} 
\to 0 \;\mbox{as}\; n\to\infty.
\ee
Finally, since $t\in [0,T]$ is arbitrary from (\ref{bound-zn}) we further have
\be\lbl{final-bound}
\lim_{n\to\infty} \P\{ \max_{t\in [0,T]} \|z_n(t)\|_{2,n}\le C_6n^{-1/2}\}=0.
\ee 
The combination of (\ref{use-Gron}) and (\ref{final-bound}) yields
that $\|\eta_n\|_{2,n}$ tends to $0$ in probability.

Using  the definitions of $\eta_n$ and $\mathbf{u}_n$, we have  
\be\lbl{final-traingle}
\|\mathbf{u}_n-\mathbf{u}\|_{C(0,T;L^2(I))}\le \max_{t\in[0,T]}\|\eta_n(t)\|_{2,n}+
\|\mathbf{v}_n-\mathbf{u}\|_{C(0,T;L^2(I))}.
\ee
Using Lemma~\ref{lem.recall} and (\ref{final-bound}), we show that 
$\|\mathbf{u}_n-\mathbf{u}\|_{C(0,T;L^2(I))}$ tends to $0$ in probability as $n\to\infty$.
\;
$\qed$

\section{Dynamical models on  W-small-world graphs}
\lbl{sec.small}
\setcounter{equation}{0}
The method developed in the previous sections can be used to derive
continuum limits for a large class of dynamical systems on random graphs.
As an application, in this section, we consider dynamical systems on SW
graphs \cite{WatStr98}. The latter are popular in modeling  networks 
of diverse nature, because they exhibit the combination of properties that are
charecteristic to both regular and random graphs, just as seen in many real-life
systems \cite{WatStr98}.

First, we introduce a convenient generalization of a SW graph. 
To this end, let $X_n$ be a set of $n$ points from $I$ as defined in (\ref{set-Xn}) and let
$W\in\mathcal{W}_0$ be a $\{0,1\}-$valued graphon. We assume that $W$ is almost everywhere
continuous on $I^2$ and its support has a positive Lebesgue measure.
Next,  define
\be\lbl{Wp}
W_p(x,y)=(1-p)W(x,y)+p(1-W(x,y)),\; p\in [0, 0.5].
\ee

\begin{df}\lbl{df.WSW} $G_n=\mathbb{G}(W_p, X_n)$ is called a $W$-small-world (W-SW) graph.
\end{df}
\begin{rem} \lbl{rem.interpol}
Note that for $p=0.5$, $W_p$ 
becomes the Erd\H{o}s-R\'{e}nyi graph $G(n,0.5)$. 
\end{rem}
\begin{rem}\lbl{rem.Wrandom}
Using the random set of points from $\tilde X_n$  as in (\ref{set-tildeXn}), one constructs 
a W-SW graph $\tilde G_n=\mathbb{G}_n(W,\tilde X_n)$ generated by 
a random set of points.
\end{rem}  
\begin{rem}
 Equation  (\ref{Wp}) implies that in the process of construction of the W-SW graph 
$G_n=\mathbb{G}(W_p, X_n)$, the new random edges to be added 
to the deterministic graph   $\mathbb{G}(W_p, X_n)$  are selected from the complement 
of the edge set $E(\mathbb{G}(W,X_n))$.
It is easy to modify (\ref{Wp}) to imitate other possible variants of the SW model.
For instance, for  fixed $q\in (0,1)$,
\be\lbl{possible}
\textbf{A)}\quad W_p=(1-p)W+pq \qquad\mbox{and}\qquad \textbf{B)}~ W_p=W+pq,\; p\in [0,1]
\ee
match the descriptions of the SW networks in \cite{WatStr98} and \cite{Mon99, NewWat99}
respectively.
\end{rem}

Theorem~\ref{thm.Wdet} shows that the continuous model (\ref{Wcont}) with $W:=W_p$ 
approximates the discrete network (\ref{Wheat}) on the W-SW graph $G_{n,p}$ for
large $n$, i.e., Equation (\ref{Wcont}) with $W=W_p$ is the continuum limit of the
discrete heat equation on the SW graph. We illustrate this result with the continuum limit for the Kuramoto model
on the SW network \cite{WilStr06,GirHas12}.

\begin{ex}\lbl{ex.Kur}
The Kuramoto model of  coupled identical phase oscillators on the SW graph $G_{n,p}$ has the following
form (cf. \cite{WilStr06})
\be\lbl{Kur}
{d\over dt} u_{ni}(t) =\omega+  \sum_{j: (i,j)\in E( G_{n,p})} \sin (2\pi(u_{nj}-u_{ni})),\; i\in [n],
\ee
where for fixed $n\in \N$ and $i\in [n]$, $u_{ni}:~\R\to \SS/\Z$ is interpreted as the phase of oscillator
$i$ and $\omega$ is the intrinsic frequency of an individual oscillator.

For this example, let
$$
X_n=\left\{0, {1\over n}, {2\over n},\dots, {n-1\over n}\right\}
$$
and 
$$
W(x,y)=\left\{\begin{array}{ll} 1,& d(x,y)\le r,\\ 0, &\mbox{otherwise}, \end{array}\right.
$$
where $d(x,y)=\min\{ |x-y|, 1-|x-y|\}$ and parameter $r\in (0,1)$ is fixed.

With the above definitions, $G_{n,p}$ is a W-SW graph. In particular, $G_{n,0}$ is 
the $k-$nearest-neighbor graph ($k=\lfloor rn\rfloor$) (see Fig.~\ref{f.2}a),
 which was used as the underlying
deterministic graph in \cite{WatStr98}, and $G_{n,0.5}$ is the Erd\H{o}sh-R\'{e}nyi graph 
$G(n,0.5)$ (see Fig.~\ref{f.2}c). Thus, the family $\{G_{n,p}\}$
interpolates between the $k-$nearest-neighbor graph and the Erd\H{o}s-R\'{e}nyi graph.
Furthermore, had we chosen to use (\ref{possible}~\textbf{A}) instead of (\ref{Wp}), we would
have obtained a family of random graph that differs from the original Watts-Strogatz SW
model \cite{WatStr98} only in minor details.

Theorem~\ref{thm.Wdet} justifies the following continuum limit for the Kuramoto model on the W-SW graph
\be\lbl{Kur-cont}
{\p\over \p t} u(x,t)=\omega+ \int_I W_p(x,y) \sin (2\pi(u(y,t)-u(x,t)))dy.
\ee
Equation (\ref{Kur-cont}) can be used to study the stability of $q-$twisted states, a family of steady state  
solutions of (\ref{Kur}), just as was done for the $k-$nearest-neighbor coupled networks in 
\cite{WilStr06,  GirHas12}.
The analysis of this problem is beyond the scope of this paper and will be presented elsewhere \cite{Med13a}. 
\end{ex}

\section{Discussion}
\lbl{sec.final}
\setcounter{equation}{0}
Coupled dynamical systems on graphs arise in modeling diverse phenomena in physics,
biology, and technology \cite{StrSync, Kur84-book, MZ12, LiErn92, PhiZan93,DorBul12, Med12}. 
The dynamics of these models is shaped by the properties
of the local dynamical systems at the nodes of the graph and the patterns of connections between
them. The principal challenge of the mathematical theory of dynamical networks is to elucidate 
the contribution of the structural properties of the networks to their dynamics. Thus,
it is important to develop analytical techniques, which apply to large classes of networks and 
reveal the interplay between the local dynamics and network topology. 
For nonlocally coupled dynamical systems, an important (albeit often formal) approach to the analysis
of network dynamics has been replacing a discrete model on a large graph with a continuum 
(thermodynamic) limit. For networks with nonlinear
diffusive coupling the continuum limit is an evolution equation with a nonlocal integral operator
modeling nonlinear diffusion. This approach has proved very useful for the 
analysis of nonlocally coupled dynamical systems on deterministic graphs 
\cite{KurBat02,AbrStr06,WilStr06,GirHas12}.

In applications, one often encounters dynamical networks on random graphs. They
are especially important in biology. For example,  random graphs are 
frequently used in computational modeling of neuronal systems, because random 
connectivity is often consistent with experimental data. For dynamical networks
on random graphs, such as SW graphs, even formal continuum limit  is
not obvious.  On the other hand, the theory of graph limits provides many examples
of convergent sequences of random graphs with relatively simple deterministic limits 
\cite{LovGraphLim12, LovSze06, BorChay11}. In \cite{GKVM13}, we used  the ideas of the 
theory of graph limits to provide a rigorous mathematical justification for 
taking the continuum limit in a large class of deterministic networks. In this paper,
we show how to derive the limiting equations for  dynamical networks on
random graphs.
Specifically, we studied coupled dynamical systems on convergent families of
$W$-random graphs \cite{LovGraphLim12,LovSze06}. The latter provide a convenient
analytical framework for modeling random graphs, which include many important examples
arising in applications,
such as Erd\H{o}s-R\'{e}nyi and SW graphs. We prove that
the solutions of the IVPs for discrete models converge in $C(0,T; L^2[0,1])$ norm 
to their continuous counterpart as the graph size goes to infinity. 

We studied networks for two variants of $W$-random graphs: those generated by the 
random and deterministic sequences respectively. The continuum limit for a family of $W$-random
graphs generated by a random sequence can be formally derived using the Monte-Carlo
method. The discrete models on graphs of this type  in general
exhibit faster convergence to the continuum limit  compared to the models of the second
type. However, the latter are more convenient in applications, as they often can 
be readily related to the existing random graph models. For example, the classical SW
graph can be interpreted as a $W$-random graph generated by a deterministic sequence.
In Section~\ref{sec.small}, we use this fact to drive the continuum limit for dynamical systems
on SW networks as an illustration of our method.  
We believe that the continuum limit analyzed in this paper will become a useful tool for studying
coupled dynamical systems on random graphs.

\vskip 0.2cm
\noindent
{\bf Acknowledgements.} 
This work was supported in part by the NSF grant  DMS 1109367.
 
\vfill\newpage
\bibliographystyle{amsplain}
\bibliography{nonlocal1}
\end{document}